\documentclass[conference]{IEEEtran}
\IEEEoverridecommandlockouts
\usepackage{cite}
\usepackage{amsmath,amssymb,amsfonts}
\usepackage{algorithmic}
\usepackage{graphicx}
\usepackage{textcomp}
\usepackage{xcolor}
\usepackage{lipsum}
\usepackage{float}
\usepackage{booktabs}
\usepackage{fancyhdr}

\makeatletter
\def\ps@IEEEtitlepagestyle{
  \def\@oddfoot{\mycopyrightnotice}
  \def\@evenfoot{}
}
\def\mycopyrightnotice{
  {\footnotesize
  \begin{minipage}{\textwidth}
  \centering
  978-1-7281-9266-6/21/\$31.00 \copyright2021 European Union.
  \end{minipage}
  }
}

\def\BibTeX{{\rm B\kern-.05em{\sc i\kern-.025em b}\kern-.08em
    T\kern-.1667em\lower.7ex\hbox{E}\kern-.125emX}}
    
\begin{document}

\title{Anomaly Detection in Cyber-Physical Systems: Reconstruction of a Prediction Error Feature Space}

\author{\IEEEauthorblockN{Nuno Oliveira}
\IEEEauthorblockA{Research Group on Intelligent Engineering and Computing \\ for Advanced Innovation and Development (GECAD) \\ Porto School of Engineering (ISEP)\\
Porto, Portugal \\
0000-0002-5030-7751}
\and
\IEEEauthorblockN{Norberto Sousa}
\IEEEauthorblockA{Research Group on Intelligent Engineering and Computing \\ for Advanced Innovation and Development (GECAD) \\ Porto School of Engineering (ISEP)\\
Porto, Portugal \\
0000-0003-2919-4817}
\and
\IEEEauthorblockN{Jorge Oliveira}
\IEEEauthorblockA{Research Group on Intelligent Engineering and Computing \\ for Advanced Innovation and Development (GECAD) \\ Porto School of Engineering (ISEP)\\
Porto, Portugal \\
0000-0002-3190-8367}
\and
\IEEEauthorblockN{Isabel Praça}
\IEEEauthorblockA{Research Group on Intelligent Engineering and Computing \\ for Advanced Innovation and Development (GECAD) \\ Porto School of Engineering (ISEP)\\
Porto, Portugal \\
0000-0002-2519-9859}
}

\maketitle

\begin{abstract}
Cyber-physical systems are infrastructures that use digital information such as network communications and sensor readings to control entities in the physical world. Many cyber-physical systems in airports, hospitals and nuclear power plants are regarded as critical infrastructures since a disruption of its normal functionality can result in negative consequences for the society. In the last few years, some security solutions for cyber-physical systems based on artificial intelligence have been proposed. Nevertheless, knowledge domain is required to properly setup and train artificial intelligence algorithms. Our work proposes a novel anomaly detection framework based on error space reconstruction, where genetic algorithms are used to perform hyperparameter optimization of machine learning methods. The proposed method achieved an F1-score of 87.89\% in the SWaT dataset.
\end{abstract}


\section{Introduction}
Cyber-physical systems (CPS) employ software components to control physical mechanisms, from simple sensors to more complex mechanical actuators and controllers \cite{baheti2011cyber,7013187}. These systems are highly dependent on network connectivity in order to be properly monitored and controlled \cite{SWaT}. The CPS opens new avenues for technological improvement. Examples of this improvement include Smart Grids \cite{SmartGrid} and Industry 4.0 \cite{industry4.0}, with real world examples such as MIT's Robot Garden \cite{RobotGarden} and CarTel \cite{CarTel}.

Due to the flexibility and the optimization afforded by CPS, the advent of smart critical infrastructures is no surprise. The examples mentioned above are but a fraction of the coming industrial revolution \cite{NextRevolution}. On the other hand, attacks can cause malfunctions of a given infrastructure or service, thus creating a negative impact in the society. Some examples have already been reported, such as the 2015 attack on the Ukrainian power grid \cite{UkrainePowerGrid}, as well as the 2006 attack that affected the water treatment plant of  Pennsylvania and the 2003 attack on a power plant in Ohio \cite{otheratt}. 

The traditional security strategies are not enough to safeguard CPS \cite{no1, no2}. The need to constantly adapt security systems to the most relevant threats makes the involvement of artificial intelligence a good solution \cite{wirkuttis2017artificial}. The applications of AI for this purpose are already present in the literature, with the application and comparison of several machine learning techniques on a popular, real world example of a cyber-physical system dataset - the Secure Water Treatment (SWaT) dataset \cite{SWaT, perales2020madics}. 

Despite the previous methodologies achieving considerable results, our novel approach of reconstruction and analysis of a new feature space has proven to be a promising research topic in the context CPS security. A Convolutional Neural Network (CNN) based predictor starts by forecasting the next expected sensor and actuator readings, when compared to the actual sensor values an error is measured at different time lags. Using that information, a new higher-dimensional space is recreated, where the system dynamics is analyzed over time. On such a space, unsupervised learning algorithms such as Support Vector Machine (SVM) and K-means clustering are used to detect deviations from normality.

The document is organized in multiple sections that can be detailed as follows. Section 2 describes the status of current security measures developed for cyber-physical systems and provides a comparison between different types of machine learning models for outlier detection. Section 3 explains the proposed anomaly detection method. Section 4 describes the considered case study. Section 5 presents the obtained results and their discussion. Section 6 provides a summary of the main conclusions that can be drawn from our research.

\section{Related Work}
The use of artificial intelligence for the security of cyber-physical systems is not a new endeavor, with several other works also presenting anomaly detection methods \cite{inoue2017anomaly, kravchik2018detecting, shalyga2018anomaly}.

In \cite{shalyga2018anomaly}, Dmitri Shalyga \textit{et al.} compared the abnormal detection rate of different types of neural networks, for identifying attacks on a CPS. The future sensor readings were predicted based on a previous window of a fixed duration. An anomaly is detected if the expected error between predicted and actual values is greater than the $99^{th}$ percentile. Furthermore, with the intention of improving the quality of the anomaly threshold, several techniques were employed such as exponentially weighted smoothing, mean p-powered error and weighted p-powered error. Finally, due to the unbalanced nature of the dataset, a disjointed time window technique was applied, where the window used to forecast at given time point might not be the one immediately preceding it. With this setup, a multilayer perceptron (MLP) based model achieved an F1-score of 81.2\%.

In \cite{kravchik2018detecting}, Moshe Kravchik \textit{et al.} used a 1D-CNN to identify anomalies in the SWaT dataset. A threshold is estimated from the model's prediction error in the training dataset. Using such a threshold, the authors were able to distinguish attacks from normal behaviours, through a z-score function. To approximate and select the local optimal hyperparameters, a grid search was employed. In this manner an F1-score of 87.1\% was achieved. In \cite{kravchik2021efficient} the same author applied the previous approach for anomaly detection, using a mix of grid search and genetic algorithms for  tuning the hyperparameters of an autoencoder network. The author obtained an F1-score of 87.3\%, an improvement from his previous work.

In \cite{zizzo2019intrusion}, Giulio Zizzo \textit{et al.}, achieved an F1-score of 81.7\% with a Long-Short Term Memory (LSTM) based model. Using grid search to compute the best values window duration and error threshold.

Jun Inoue \textit{et al.} in \cite{inoue2017anomaly} have investigated the detection capabilites of two unsupervised machine learning algorithms, a Deep Neural Network (DNN) composed of a single LSTM layer followed by multiple feed-forward layers and a one-class SVM. The first implements probabilistic outlier detection by judging low probability data points as anomalies while the second resembles a more straightfoward application of the one-class SVM. For the SWaT dataset the DNN obtained an f1-score of 80.3\% as the one-class SVM achieved a slightly smaller value for the same metric, 79.6\%. Furthermore, the authors have concluded that both methods share the same limitations: detecting anomalous actuator behaviour and identifying gradual changes of sensor values.

In \cite{Li2019MADGANMA}, D. Li \textit{et al.} presents a novel anomaly detection approach using Generative Adversarial Networks (GAN). In this work two LSTM Recurrent Neural Networks (LSTM-RNN) are trained in different roles, one is trained as a detector finding anomalies that deviate from the baseline, and the other as a generator capable of creating simulated normal baselines. These two models are then introduced as adversaries in an iterative feedback-loop training process, obtaining an F1-score result of 77\% on the SWaT testbed.

For this dataset, the best results we found in the literature, 88.2\% F1-score, was obtained by M. Elnour \textit{et al.} in \cite{elnour2020dual}. An isolation forest based approach to anomaly detection, that separates anomalies from normal observations by analysing both normal data and its Principal Component Analysis (PCA) transformed representations.

\section{Proposed Method}

\begin{figure*}[t]
\centering
\includegraphics[height=5 cm]{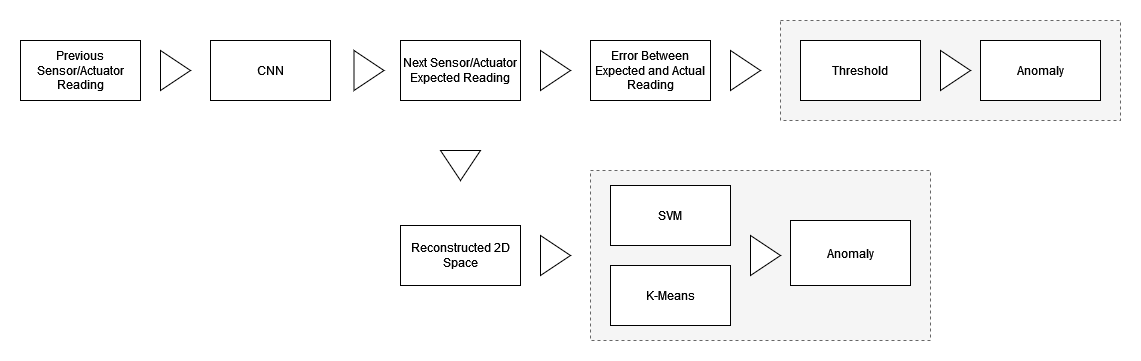}
\caption{Overview of the proposed anomaly detection approach.}
\label{fig:anomaly}
\end{figure*}

Our anomaly detection approach is based on the forecasting of sensor and actuator reading values for the near future through a machine learning model. This model was trained only on normal data, i.e. data extracted from the expected system activity. By comparing the model's predictions with the ground truth (future sensor readings) we are able to compute the model's prediction error. In the deployment phase, an attack that compromises the expected system activity, generates signatures or patterns that a machine learning model  (trained on regular data activity) is not capable of predict. When these unexpected and large prediction errors happen, the AI system can infer that an attack was performed on the system, and stop it in an early stage. Figure \ref{fig:anomaly} provides an overview of the proposed anomaly detection approach.

For the forecasting task, a CNN is used. These are very popular methods in image processing and object recognition tasks \cite{cnn}. For time-series analysis, CNNs are also widely used due to their receptive field capability, which allows them to explore efficiently the time dependencies across several events of a given time frame. On the other hand, for the anomaly identification task we have considered three different approaches, a fixed anomaly detection threshold ($\alpha$) obtained by multiplying the largest  model's prediction error in the train set ($\delta$) with a sensitivity parameter ($\beta$); one-class support vector machine and K-means clustering.

The employed deep learning model has several hyperparameters and architectures that can be explored to achieve optimal results. Furthermore, the whole anomaly detection approach also depends on other parameters such as the window, $w$, and the sensitivity parameter, $\beta$. As the readings for the utilized dataset are collected every second, this implicates a window of $w$s. Table 1 describes a base architecture for the CNN. For each convolutional layer of the presented architecture, padding was applied equally and stride was set to one.

\begin{table}[H]
\caption{Example - CNN architecture.}
\centering
\footnotesize
\begin{tabular}{llll}
\toprule
\textbf{Layer}	& \textbf{Size}  & \textbf{Activation}	& \textbf{Dropout}\\
\midrule
Input           & ($w$, 51) &   -           &   -\\
Conv1D          & 32        &   ReLU        &   -\\
MaxPooling1D    & 2	        &   -           &   -\\
Conv1D          & 64        &   ReLU        &   -\\
MaxPooling1D    & 2         &   -           &   -\\
Flatten         & 192       &   -           &   -\\
Dense           & 64        &   tanh        &   0.2\\
Dense           & 32        &   tanh        &   0.2\\
Dense           & 51        &   sigmoid     &   0.2\\
\bottomrule
\end{tabular}
\label{tab:mlparch}
\end{table}

The network was trained for 100 epochs using batches of 433 and Adam as optimization function. Additionally, an early stopping method was employed in order to stop the training as soon as the loss value stopped increasing for five consecutive epochs.

There is a vast number of hyperparameter combinations for the proposed method. Exploring all possibilities with a grid search would be impractical and reducing the amount of possible values that each hyperparameter can take would drastically reduce the chances of finding an optimal network setup. Therefore, we opted to use a genetic algorithm to incrementally search for solutions with better detection performance through a series of reproduction, mutation, evaluation and selection processes. The hyperparameters of the proposed detection method were encoded as genes of individuals which resemble candidate solutions.

Regarding the unsupervised learning models, a new embedding space was created from the CNN forecasting errors and their lagged observation values. The SVM was directly trained on the errors of the CNN train set forecasts to build a decision function able to distinguish errors related to normal behaviour from abnormal-related ones. Radial basis function was used as kernel function of the SVM model and different values for $\mu$ and $\gamma$ were experimented.

For the K-means, the errors of the CNN train set forecasts were insufficient to train the model since examples of errors resembling attack occurrences are necessary to make the algorithm categorize the data points into different clusters - one for normal errors and another one for attack-related errors.

Regarding the model configuration, the Euclidean distance function was used and the clusters were initialized using the K-means++ algorithm. The stopping criterion was set to 300 iterations.

\section{Case Study}

In order to evaluate the proposed method, we have selected a case study in the context of cyber-physical systems research, the SWaT dataset. SWaT was designed by the Singapore University of Technology and Design to boost CPS security research. They provided a labeled dataset collected from a  realistic testbed of sufficient complexity \cite{SWaT}. It represents a scaled-down version of real-world industrial water treatment plants such as those found in the majority of cities.

\begin{figure}[H]
\centering
\includegraphics[width=8.5 cm]{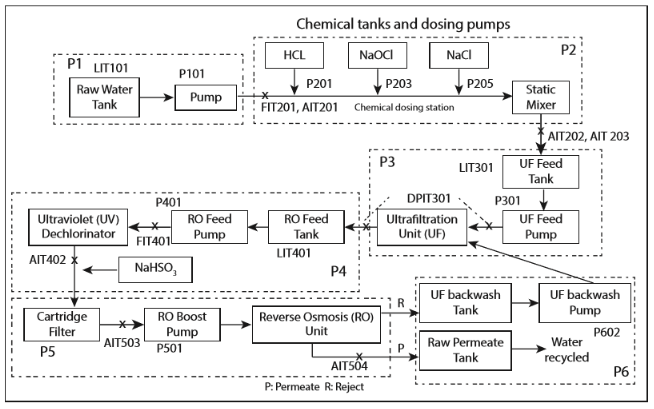}
\caption{SWaT processes overview (extracted from \cite{SWaT}) }
\label{fig:swat}
\end{figure}

Data was recorded from the sensors and actuators distributed across the 6 steps of the water treatment process, illustrated in Figure \ref{fig:swat}. The 24 sensors measure water level, flow, PH and conductivity, among other things, in each relevant step of the process. This information is then acted upon by the multitude of actuators, controlling pumps and motorized valves. 

All sensor and actuator readings were logged every second for a total of eleven days. In the first seven days only normal activities were registered, while in the last four days both normal activities and several attacks launched at the water treatment plant were recorded. The collected data concerns both physical properties related to the water treatment process and network traffic.

A total of 36 attacks were launched at the testbed and their duration may vary based on the attack type. These can be categorized into four types: Single Stage Single Point (SSSP), Single  Stage  Multi-Point (SSMP), Multi Stage Single Point (MSSP), Multi Stage Multi-Point (MSMP), for more details the reader can consult \cite{SWaT}.

In Table 2, the number of attacks per category is described.

\begin{table}[H]
\caption{Number of attacks per category \cite{SWaT}}
\centering
\begin{tabular}{lc}
\toprule
\textbf{Category} & \textbf{Nº of Attacks}	\\
\midrule
SSSP    &   26 \\ 
SSMP    &   4  \\
MSSP    &   2  \\
MSMP    &   4  \\
\bottomrule
\end{tabular}
\end{table}

Each reading was labeled as either normal or an attack. The data totals near one million entries and is presented in two separate files, one with 495000 readings of normal behavior and another with 449919 readings of which 54621 resemble attack occurrences. The dataset is very challenging for supervised learning classification algorithms since the testbed is highly unbalanced towards the minority class. Our models were trained using data from the first file, i.e. data extracted from the normal system activity. In the testing phase, the second file is used instead, where the model's ability to identify unusual and abnormal behaviours in the physical  system is measured.

In order feed the CNN, the SWaT data was first preprocessed. The data was indexed to ensure the correct ordering of the series. Furthermore, the 51 sensor readings were normalized using min-max method.

\subsection{Data Augmentation}
Although the SWaT dataset is composed by thousands of records from past attacks, it's train set file lacks on extreme sensor and actuator readings that are often registered during an attack. Hence, the errors of the CNN predictor, which was trained only on normal behaviour, are also very small and unrepresentative of those related to malicious attempts. Algorithms like the K-means which operate in the new embedding space recreated from the CNN prediction errors require these extreme values to build an efficient decision boundary. Therefore, and in an attempt to recreate such attack patterns, synthetic data sampled from a two dimensional Gaussian probability density distribution was added to the training set of the new embedding space.
The parameters of the Gaussian probability density distribution were inferred from the analysis of the train set. $148497$ data points were generated (30\% of the train set size) with a mean vector $[ 2 \cdot \delta, 2 \cdot \delta]$. The diagonal elements of a $2\times 2$ co-variance matrix are $diag(\Sigma) = [\sigma_{train}, \sigma_{train}]$, off-diagonal elements are equal to zero. The  $\sigma_{train}$ is the standard deviation of the CNN prediction errors in the train set.

\subsection{Evaluation Metrics}

In order to accurately forecast the readings of the physical devices for the next time step, the CNN was trained to minimize the mean average error (MAE) between its prediction and the ground truth. 



The MAE was also used to determine the value of the anomaly threshold, $\alpha$. This threshold works as a decision boundary, thus separating the space between normal events and attack. Usually, accuracy is one of the most common metrics to evaluate the performance of a classifier, however, it is biased towards the majority class. Our work, similarly to others in literature \cite{cnn, autoencoder}, adopted the F1-score as an evaluation metric due to the highly robustness to unbiased dataset, for more details the reader can be redirected to \cite{app11041674}.

\section{Results and Discussion}

The experimental settings for the presented results were a single machine running Windows Server 2012 as Operating System, equipped with 64 gigabytes of RAM, approximately 4 terabytes of disk space, an Intel(R) Xeon(R) CPU with 24 cores and 4 NVIDIA Tesla k20c GPUs.

\subsection{Anomaly Detection Threshold}

\begin{figure}[H]
\centering
\includegraphics[height=5 cm]{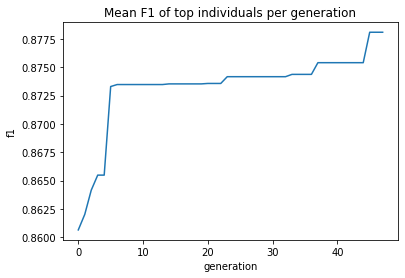}
\caption{Mean of the F1-score value of the top individuals.}
\label{fig:mean}
\end{figure}

The proposed genetic algorithm was used to optimize the hyperparameters of the approach which combines the CNN and the fixed anomaly detection threshold so that the value of F1-score could be maximized. Figure \ref{fig:mean} describes how the F1-score of the best individuals of the genetic algorithm evolved over 47 iterations.

It can be observed that there was a very significant increase of the F1-score value over the first 5 generations, from $86.06\%$ to $87.33\%$, and then that value steadily increased over the course of the remaining generations up to $87.81\%$. The results obtained for the best two individuals are described in Table 3.

\begin{table}[H]
\caption{Overall results of the best two individuals.}
\centering
\begin{tabular}{ccccc}
\toprule
\textbf{Individual} & \textbf{Accuracy}	& \textbf{Precision}  & \textbf{Recall}	& \textbf{F1-score}\\
\midrule
Best            &   95.47   &   94.54   &   83.38   &   87.89   \\ 
Second Best     &   95.57   &   96.94   &   82.14   &   87.72   \\
\bottomrule
\end{tabular}
\end{table}

The best individual has the highest F1-score value, $87,89\%$, since its precision and recall values are better balanced than the ones presented by the second best individual, which has greater precision but a lower recall.

\subsection{Unsupervised Methods}

The anomaly detection threshold, although it suggests a reasonable decision boundary, it is highly dependent on outliers in the training dataset. In order to solve this issue, we employed two unsupervised methods, one-class SVM and K-means clustering which are able to learn an effective decision boundary function directly from data. Thereby, with the prediction errors we elaborated a new observation space composed of two features, the prediction error of each observation and its lagged value. This approach enabled us to construct a two-dimensional space on which we applied both methods.

With the one-class SVM, we used the prediction errors of the train data in order to infer/predict and attack. Initially, we obtained a poor decision function, F1-scores around $60\%$. Later on, we decided to weight each sample proportionally to its error value in order to favour outliers. This resulted in a major improvement in the F1-score value, reaching up to $74,76\%$. The results of the one-class SVM model are presented under Table 4.

Although the results are not as good as the ones presented by threshold-based anomaly detection algorithm we believe that the reconstruction of a more interesting embedding space where attacks are more easily detected is an important research line.

Regarding the K-means clustering algorithm, we used the errors of the train set and synthetic samples generated based on a Gaussian distribution to fit the algorithm in order to obtain a segregation between two separate groups, normal and attack. This approach worked better than the one-class SVM, obtaining an F1-score of $87.16\%$. The result of the K-means model is presented in Table 4.

\begin{table}[H]
\caption{Results for the unsupervised models.}
\centering
\begin{tabular}{ccccc}
\toprule
\textbf{Model} & \textbf{Accuracy}	& \textbf{Precision}  & \textbf{Recall}	& \textbf{F1-score}\\
\midrule
K-Means   &   95.41   &   96.98   &   81.40   &   87.16   \\
One-class SVM   &   87.11   &   71.79   &   80.00   &   74.76   \\ 
\bottomrule
\end{tabular}
\end{table}

\section{Conclusion}

We have presented a novel anomaly detection method based on machine learning models in which the forecasting algorithm is automatically optimized by a genetic algorithm. The proposed framework uses a CNN to forecast sensor and actuator readings and utilizes the errors obtained from the difference between the predicted values and the ground truth to measure deviations from the normal system behavior. For this task, we compared the performance of three different methods, an engineered threshold based on the train set errors, an SVM and K-means clustering. 

With our approach we have achieved an F1-score of $87.89\%$, in line with the current state-of-art results. As future work, embedding spaces are going to be explored more efficiently.

\section*{Acknowledgments}
The present work was done and funded in the scope of European Union's Horizon 2020 research and innovation programme under project SeCoIIA (grant agreement No 871967). This work has also received funding from projects UIDB/00760/2020 and UIDP/00760/2020.

\bibliographystyle{ieeetr}
\bibliography{bibliography}

\end{document}